\documentclass[reprint,aps,epsfig,superscriptaddress,twocolumn]{revtex4-1}
\usepackage{bm}
\usepackage{times}
\usepackage[dvips]{graphicx}
\usepackage{mathrsfs}
\usepackage[intlimits]{amsmath}
\usepackage{textcomp}
\usepackage[colorlinks, citecolor=red]{hyperref}
\usepackage{tabularx}
\usepackage{xcolor}
\usepackage{color}
\usepackage[defaultcolr=black]{changes}
\usepackage{ulem}
\usepackage[switch]{lineno}
\usepackage{setspace}

\begin{document}

\title{A metropolitan-scale trapped-ion quantum network node with hybrid multiplexing enhancements}

\author{Z.-B. Cui}
\affiliation{Center for Quantum Information, Institute for Interdisciplinary Information Sciences, Tsinghua University, Beijing 100084, PR China}

\author{Z.-Q. Wang}
\affiliation{Center for Quantum Information, Institute for Interdisciplinary Information Sciences, Tsinghua University, Beijing 100084, PR China}

\author{P.-C. Lai}
\affiliation{Center for Quantum Information, Institute for Interdisciplinary Information Sciences, Tsinghua University, Beijing 100084, PR China}

\author{Y. Wang}
\affiliation{HYQ Co., Ltd., Beijing 100176, PR China}

\author{J.-X. Shi}
\affiliation{Center for Quantum Information, Institute for Interdisciplinary Information Sciences, Tsinghua University, Beijing 100084, PR China}

\author{P.-Y. Liu}
\affiliation{Center for Quantum Information, Institute for Interdisciplinary Information Sciences, Tsinghua University, Beijing 100084, PR China}

\author{Y.-D. Sun}
\affiliation{Center for Quantum Information, Institute for Interdisciplinary Information Sciences, Tsinghua University, Beijing 100084, PR China}

\author{Z.-C. Tian}
\affiliation{Center for Quantum Information, Institute for Interdisciplinary Information Sciences, Tsinghua University, Beijing 100084, PR China}

\author{Y.-B. Liang}
\affiliation{Center for Quantum Information, Institute for Interdisciplinary Information Sciences, Tsinghua University, Beijing 100084, PR China}

\author{B.-X. Qi}
\affiliation{Center for Quantum Information, Institute for Interdisciplinary Information Sciences, Tsinghua University, Beijing 100084, PR China}

\author{Y.-Y. Huang}
\affiliation{Center for Quantum Information, Institute for Interdisciplinary Information Sciences, Tsinghua University, Beijing 100084, PR China}

\author{Z.-C. Zhou}
\affiliation{Center for Quantum Information, Institute for Interdisciplinary Information Sciences, Tsinghua University, Beijing 100084, PR China}
\affiliation{Hefei National Laboratory, Hefei 230088, PR China}

\author{Y.-K. Wu}
\affiliation{Center for Quantum Information, Institute for Interdisciplinary Information Sciences, Tsinghua University, Beijing 100084, PR China}
\affiliation{Hefei National Laboratory, Hefei 230088, PR China}

\author{Y. Xu}
\affiliation{Center for Quantum Information, Institute for Interdisciplinary Information Sciences, Tsinghua University, Beijing 100084, PR China}
\affiliation{Hefei National Laboratory, Hefei 230088, PR China}

\author{L.-M. Duan}
\email{lmduan@tsinghua.edu.cn}
\affiliation{Center for Quantum Information, Institute for Interdisciplinary Information Sciences, Tsinghua University, Beijing 100084, PR China}
\affiliation{Hefei National Laboratory, Hefei 230088, PR China}

\author{Y.-F. Pu}
\email{puyf@tsinghua.edu.cn}
\affiliation{Center for Quantum Information, Institute for Interdisciplinary Information Sciences, Tsinghua University, Beijing 100084, PR China}
\affiliation{Hefei National Laboratory, Hefei 230088, PR China}

\begin{abstract}
Quantum network and quantum repeater are promising ways to scale up a quantum information system to enable various applications with unprecedented performance. As a current bottleneck of building a long-distance quantum network, the distribution rate of heralded entanglement between remote network nodes is typically much lower than the decoherence rate of each local node, which obstructs the implementation of a metropolitan-scale quantum network with more than two remote nodes. A promising scheme to accelerate the remote entanglement distribution is through multiplexing enhancement based on a multimode quantum network node. In this work, we experimentally realize a functional $5$-ion quantum network node with two different types of qubits inside. We employ a hybrid multiplexing scheme combining the methods of multiple excitation and ion shuttling, in which maximally $44$ time-bin modes are generated and sent through a long fiber to boost the entangling rate. Via this scheme, we can generate heralded ion-photon entanglement with a high fidelity of $96.8\%$/$94.6\%$/$89.8\%$ with a success rate of $263\,\text{s}^{-1}$/$40\,\text{s}^{-1}$/$4.28\,\text{s}^{-1}$, over a fiber of $3\,$m/$1\,$km/$12\,$km, respectively. In addition, the memory qubit can protect the stored quantum information from the destructive ion-photon entangling attempts via dual-type encoding and a memory coherence time of $366\,$ms is achieved. This coherence time has exceeded the expected entanglement generation time $234\,$ms over a $12\,$km fiber, which is realized for the first time in a metropolitan-scale quantum network node.
\end{abstract}

\maketitle

\section{Introduction}

Quantum network and quantum repeater are among the most promising ways to extend the size of a quantum system, which can enable important applications such as distributed quantum computing~\cite{distributed, duan arxiv, duan pra,oxford distributed}, long distance quantum communication~\cite{BDCZ,DLCZ,rmp_gisin,oxford_qkd,weifurner_qkd} and network-enhanced quantum metrology~\cite{ye_and_lukin,repeater_telescope, oxford_clock}. In a quantum network or quantum repeater protocol, one of the core tasks is distributing heralded quantum entanglement between distant nodes. So far, heralded entanglement between two quantum network nodes has been demonstrated experimentally over either short (lab size) or long (metropolitan size) distances~\cite{oxford distributed,monroe_barium,sc_bell_inequality,solid ion,rempe,230m,bao3nodes,weinfurter_33km,2021nature,hanson long,lukin reflection}. Currently quantum network has been realized with various physical systems. Among all these physical platforms, trapped atomic ion is a promising candidate. High local gate fidelity (better than 99.9\%~\cite{quantiuum,oxford_gate}, critical to entanglement swapping and entanglement distillation), long coherence time ($\sim1$ hour~\cite{1hour}), and record-high rate ($\sim200\,$Hz~\cite{monroe_barium}) and fidelity ($\sim96\%$~\cite{oxford distributed}) between two network nodes have been demonstrated separately in this system.

Despite all these advances achieved so far, currently it is still very difficult to demonstrate a quantum repeater with three nodes or a quantum network which can utilize more than one pair of remote entanglement, when the separation between different nodes reaches metropolitan scale ($\sim10\,$km). A requisite for this outstanding goal is the rate of remote entanglement generation between different nodes should be higher than the memory decoherence rate of each network node, as the remote entangling operation is in a probabilistic way~\cite{BDCZ,DLCZ}. This stringent requirement can also be interpreted as the `link efficiency' $\eta_{\,\text{link}}$ should be larger than unity~\cite{hansondelivery,duan pra}. Currently, $\eta_{\,\text{link}}>1$ can only be achieved in a lab-scale quantum network~\cite{hansondelivery,monroe}. For all the heralded entanglement between two nodes over a metropolitan-scale fiber ($>10\,$km), the link efficiency is still less than $0.01$.

One of the most important obstacles which limits the link efficiency for a metropolitan-scale quantum network lies in the round-trip heralding time in each remote entangling attempt. For a $10\,$km fiber, the time cost for each entangling attempt is at least $2L/c=100\,\mu$s ($L$ is the fiber length and $c$ is the light speed), which is two orders of magnitude higher than the value in a lab-scale case~\cite{oxford distributed,monroe}. To resolve this issue, multimode quantum repeater~\cite{2007prl,rmp_gisin} has been proposed, which is able to accelerate the remote entangling operation via a multiplexing scheme. Currently, multiplexed quantum network node has been demonstrated in several different physical systems for quantum memories ~\cite{2021nature,1250,tittel_frequency,oam,lan,225,multipurpose,enhancement,changwei,wavevector,lanyon multiplexing,haffner multiplexing, solid ion}. Multiplexing-enhanced heralded atom-photon entanglement over a metropolitan-scale fiber has been demonstrated on neutral atomic ensemble system recently, and a link efficiency as high as $0.46$ has been achieved~\cite{enhancement}.

\begin{figure*}[]
    \centering
    \includegraphics[width=1.0\textwidth]{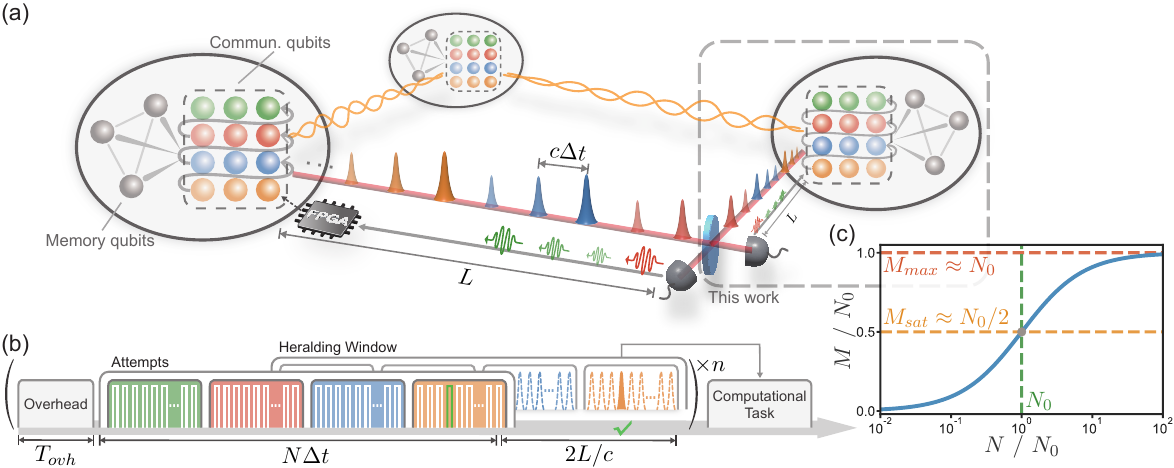}
    \caption{
    A multiplexed quantum network.
    (a) The schematic of a multiplexed trapped-ion quantum network. There are two kinds of qubits in each network node. The communication qubits (green, red, blue, orange) are used for distributing heralded entanglement between different nodes, and the memory qubits (grey) are responsible for storing local quantum information or carrying out local computing tasks. Entanglement between distant nodes is heralded via interfering the photons on a beamsplitter in the middle point of the two nodes. The entanglement distribution task can be accelerated via different multiplexing schemes.
    (b) The protocol of a time-bin multiplexing scheme. Totally $N$ temporal modes are excited in one run of entangling attempts and the corresponding heralding signal returns to the node after a round-trip travel time of $\frac{2L}{c}$ following the excitation.
    (c) Enhancement factor in time-bin multiplexing. The enhancement factor $M$ starts saturating when the duty cycle of the entangling attempts reaches $50\%$. Here $N_0=(\frac{2L}{c}+T_{\text{ovh}})/\Delta t$ is the characteristic number when the duty cycle reaches $50\%$.
    }
    \label{FIG1}
\end{figure*}

In this work, we experimentally realize a highly multiplexed trapped-ion quantum network node which can distribute heralded ion-photon entanglement  with a high rate and protect the quantum information stored locally. We demonstrate different multiplexing schemes for three different fiber lengths at $3\,$m (lab size), $1\,$km (campus size), and $12\,$km (metropolitan size). The heralded ion-photon entanglement distribution rate is improved by $3$-fold, $5$-fold, and $16$-fold in these three cases respectively.  For the $12\,$km case, a hybrid scheme combing both the multiple excitation and ion shuttling is employed to boost the remote ion-photon entangling rate to $4.28\, \text{s}^{-1}$, which is already higher than the rate of memory decoherence $3.69\,$s$^{-1}$ and yields a record-high link efficiency of $1.16$ for ion-photon entanglement over a metropolitan-scale fiber. Besides, with the dual-type encoding~\cite{huangyy,lai,yanghx,omg}, the local quantum information stored in the memory qubit is not influenced by the noisy operations (such as cooling, pumping, and photon excitation) during the ion-photon entangling attempts, which is also a requisite for building a multi-node ($>2$) quantum network and quantum repeater. In addition, the spontaneous emission error in the memory qubit can be detected by a mid-circuit measurement, demonstrating another important capability regarding the purification and error correction needed in a future large-scale quantum network~\cite{duan pra}. With all these achievements, this functional and highly-multiplexed trapped-ion quantum network node will contribute to the realization of a multi-node metropolitan-scale quantum network (Fig.~\hyperref[FIG1]{1(a)}) in the future .

\section{Results}
\subsection{Enhancement factor of time-bin multiplexing}\label{text:efc}

There are already several different multiplexing schemes demonstrated so far, which exploit the time-bin, frequency, orbital angular momentum (OAM), or other dimensions of a flying qubit. A multiplexing scheme effectively reduces the averaged time cost of each entangling attempt, thus accelerates the entanglement distribution process. For a time-bin multiplexing scheme shown in Fig.~\hyperref[FIG1]{1(b)}, the averaged time cost for each entangling attempt can be expressed as:
\begin{equation}\label{Teff}
T_{\text{eff}}=(\frac{2L}{c}+T_{\text{ovh}}+N \Delta t)/N= \frac{\frac{2L}{c}+T_{\text{ovh}}}{N}+\Delta t
\end{equation}
where $\frac{2L}{c}$ is the round-trip travel time in the long fiber ($L$ is the fiber length and $c$ is the light speed in fiber), $T_{\text{ovh}}$ is the averaged time overhead in each round of $N$ entangling attempts (including communication time in control system, initial cooling before each entangling round starts, sympathetic cooling in the entangling process, and all other overhead in the sequence), $N$ is the number of modes used in each round of entangling attempts, and $\Delta t$ is the time interval between two successive modes. Usually, we have $\frac{2L}{c}+T_{\text{ovh}}\gg \Delta t$. In the limit of a large $N$, the effective time cost for an entangling attempt $T_{\text{eff}}$ will be dominated by $\Delta t$, as shown in Eq.~\eqref{Teff}. To achieve a high effective attempting rate $1/T_{\text{eff}}$ in the process of remote entangling, we need to improve the number of time-bin modes ($N$) and select an appropriate time interval between them ($\Delta t$). It is also noteworthy that the frequency or OAM multiplexing can be considered as a special case with $\Delta t=0$ in this model.

\begin{figure*}[htbp]
    \centering
    \includegraphics[width=1.0\textwidth]{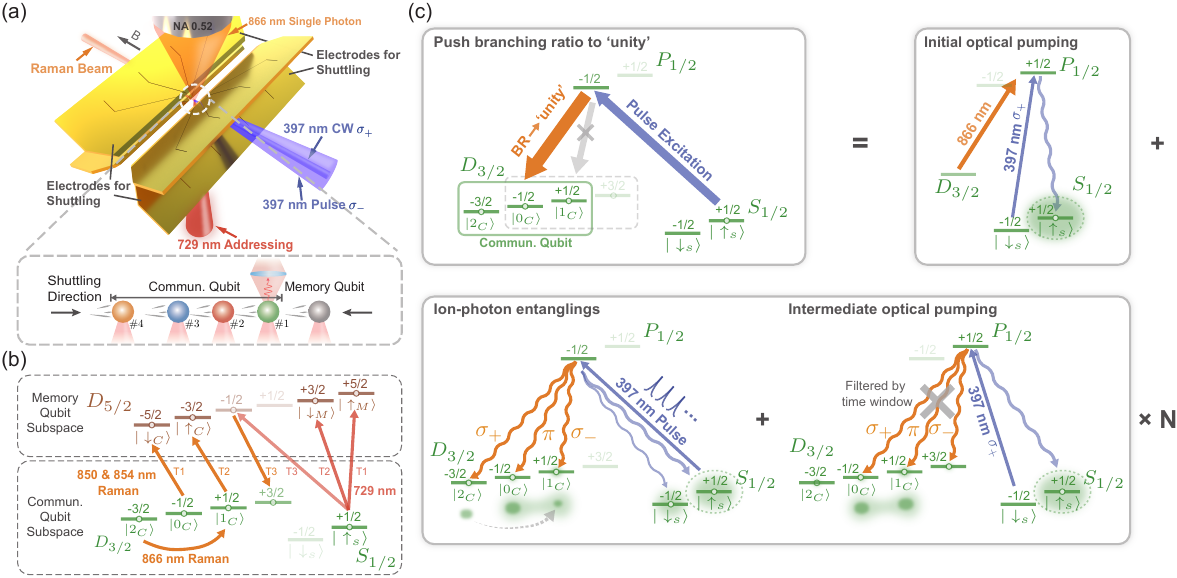}
    \caption{
    Experimental setup, encoding scheme, and multiplexing protocols.
    (a) In this experiment, a $5$-ion (including $4$ communication qubits and $1$ memory qubit) crystal is confined in a linear blade trap. The magnetic field is perpendicular to the ion chain.
    An $866\,$nm objective, aligned perpendicular to the magnetic field, collects single photons in ion photon entangling process.
    Four DC electrodes are employed to shuttle the ion chain for multiplexing.
    A $729\,$nm addressing beam is able to drive each ion individually. Other beams such as the $397\,$nm $\sigma^+$ optical pumping, $397\,$nm picosecond pulse for entangling, two sets of Raman beams, and the cooling and pumping beams (not shown in this figure) are all global.
    (b) Dual-type encoding of the communication qubits and memory qubits. The communication qubits are encoded in a closed three-level subspace consisting of $|S_{1/2}\rangle$, $|D_{3/2}\rangle$, and $|P_{1/2}\rangle$, which is able to carry out cooling, pumping, ion photon entangling, and detection without influencing the memory qubit subspace $|D_{5/2}\rangle$. Coherent conversions between these two subspaces can be realized with $729\,$nm quadruple transition or a $850\,\text{nm}/854\,\text{nm}$ Raman transition. Another $866\,\text{nm}/866\,\text{nm}$ Raman transition is used for rotation in the $|D_{3/2}\rangle$ level~\cite{lai}.
    (c) The multiplexing scheme of multiple excitation. We squeeze the population out of the $|S_{1/2}\rangle$ level by multiple ion-photon entangling pulses and this can effectively increase the branching ratio of the $866\,$nm transition from $6\%$ to `unity'. After several excitation pulses, the population accumulated in $|S_{1/2},m=-1/2\rangle$ is cleaned by an intermediate optical pumping. As the ion can also decays to $|D_{3/2}\rangle$ during the optical pumping, the optimal branching ratio we can achieve is $54.4\%$, which is close to but slightly lower than `unity'.
    }
    \label{FIG2}
\end{figure*}

Based on Eq.~\eqref{Teff}, we can derive the enhancement factor of multiplexing as:
\begin{equation}\label{M}M=T_0/T_{\text{eff}}\end{equation}
where $T_0=\frac{2L}{c}+T_{\text{ovh}}+\Delta t$ is the averaged time cost of each entangling attempt with a single-mode quantum memory. Note that here we assume the success probabilities in all the modes are the same. We illustrate the multiplexing enhancement versus mode number in Fig.~\hyperref[FIG1]{1(c)}. One can see that given a fixed time interval $\Delta t$, the multiplexing enhancement $M$ increases with $N$ and saturates around $N_0=\frac{\frac{2L}{c}+T_{\text{ovh}}}{\Delta t}$, where the duty-cycle of entangling attempts reaches $50\%$ (duration for entangling attempts $N\Delta t$ is equal to all the waiting time $\frac{2L}{c}+T_{\text{ovh}}$). In this case a multiplexing enhancement of $M=\frac{N_0+1}{2}\approx\frac{N_0}{2}$ can be achieved. This $50\%$ duty cycle can be regarded as an indication that `sufficient' multiplexing has been employed, as further increase of $N$ can only improve the multiplexing enhancement by a factor of less than twice.

In the realistic case, the success probabilities for different modes are usually not the same due to the inevitable inhomogeneity in the system. Therefore, the effective multiplexing enhancement of a multi-mode quantum network over a single-mode case needs a more complicated evaluation. However, the simplified model in Eq.~\eqref{Teff} still reveals the essentials and is able to provide a quick estimation on the effectiveness of multiplexing. More detailed calculation of multiplexing enhancement with mode inhomogeneity considered is described in Appendix \ref{app:dmec}.

\subsection{Two different multiplexing methods}\label{text:mem}

We use the dual-type scheme to build our trapped-ion quantum network node~\cite{lai,huangyy,yanghx,omg}. Following the scheme demonstrated in~\cite{lai} and Fig.~\hyperref[FIG2]{2}, we can implement the cooling, pumping, state detection, and ion-photon entanglement excitation with $397\,$nm and $866\,$nm lasers in a closed three-level communication qubit subspace consisting of $|S_{1/2}\rangle$, $|D_{3/2}\rangle$, and $|P_{1/2}\rangle$ levels of $^{40}$Ca$^{+}$. During the dissipative operations such as cooling, pumping, detection, and ion-photon entangling, the local quantum information can be faithfully stored in the memory qubit subspace ($|D_{5/2}\rangle$ level), without being influenced due to large spectral separation over THz. The coherent conversion between these two subspaces can be implemented by a $729\,$nm quadruple transition or an $850\,\text{nm}/854\,\text{nm}$ Raman transition with high fidelities~\cite{lai}.

\begin{figure*}[htbp]
    \centering
    \includegraphics[width=0.95\textwidth]{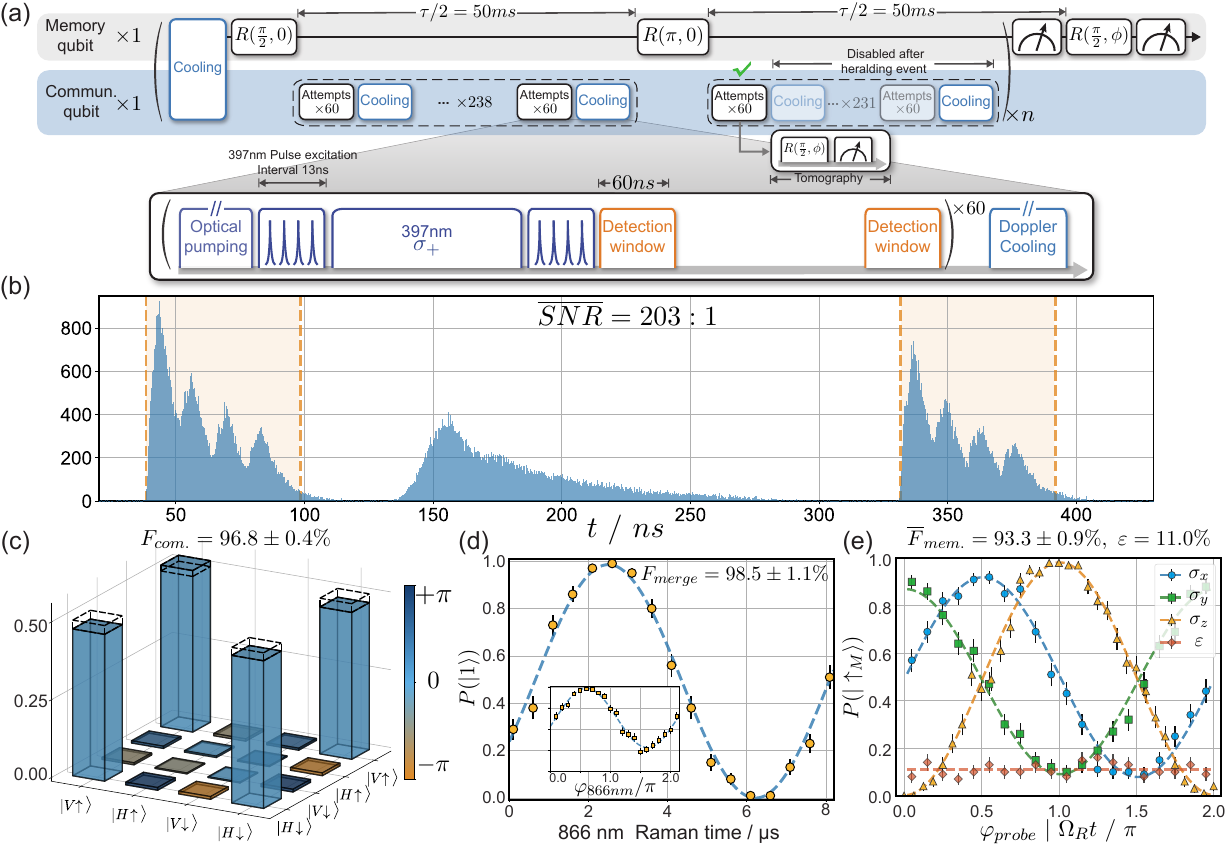}
    \caption{
        Experimental results of multiplexing-enhanced ion photon entanglement in $3\,$m case.
        (a) Experimental protocol.
        We apply a spin echo sequence of $\tau$ on the memory qubit.
        Meanwhile, the ion-photon entangling attempts are applied to the communication qubit.
        In each $\tau/2$, the blocks of entangling attempts are repeated for $238$ and $231$ times. In each block, series of $8$ ion photon entangling attempts are repeated for $60$ times. Each block is followed by a $120\,\mu$s sympathetic Doppler cooling. Each series of $8$ attempts lasts for $1.5\mu$s (including the initial optical pumping).
        Upon detecting a photon, all the following entanglement attempts are halted.
        State analysis of the communication qubit can be executed directly after the heralding event or delayed until later computational tasks with the memory qubit. The protocol restarts if no photon is detected.
        (b) The histogram of the photon detection events in the ion photon entangling attempts. The photon is in $866\,$nm. The photon detection events between $150\,$ns and $250\,$ns are induced by the intermediate optical pumping.
        (c) The reconstructed density matrix of the ion-photon entanglement. The measured fidelity is $96.8\pm0.4\%$. By applying real-time phase adjustment, we can yield the same ion-photon entangled state no matter which of the $8$ modes is excited.
        (d) Merging $|2_C\rangle$ and $|1_C\rangle$ by a $866\,$nm$/$$866\,$nm Raman pulse (see Fig.~\hyperref[FIG2]{2(b)} and \cite{lai}). A $\frac{2}{3}\pi$-pulse is first used to scan the phase for merging the levels, as indicated in the inset.
        A subsequent fine scan on time is performed at the optimal phase. This measurement yields a merging fidelity of $98.5\pm1.1\%$.
        (e) The memory qubit under different mutually unbiased bases (MUBs).
        The memory qubit is initialized to six MUBs which are also the eigenstates of $\sigma_x$, $\sigma_y$ and $\sigma_z$.
        By scanning the phase or time of the $729\,$nm pulse, we can perform the single qubit state tomography and get the fidelity of each stored qubit.
        Mid-circuit detection is employed to detect the errors $\epsilon$ caused by the finite lifetime of $|D_{5/2}\rangle$ state. If a decay event is detected in the mid-circuit measurement, this trial is aborted and the protocol starts over again.
        With a success rate of $89\%$, the memory qubit fidelity averaged over six MUBs reaches $93.3\pm 0.9\%$.
    }
    \label{FIG3}
\end{figure*}

The photon we collect in ion-photon entanglement generation is from the $|P_{1/2}\rangle$ to $|D_{3/2}\rangle$ transition in $866\,$nm wavelength, which is very suitable for being converted to telecom C band with high efficiency and low noise via the difference frequency generation (DFG) in a Periodically Poled Lithium Niobate (PPLN) waveguide. However, the small branching ratio of $6\%$ limits the efficiency of the ion-photon entanglement generation. Therefore, in the first multiplexing scheme, we demonstrate how to resolve this issue by a method of multiple excitation.

The idea is that we can reuse the ion after the unsuccessful excitation. Each time when the ion is excited to the $|P_{1/2}\rangle$ level, the ion decays back to the initial state $|S_{1/2},m=+1/2\rangle$ with a probability of roughly $\frac{2}{3}\times(1-6\%)\approx63\%$, to $|D_{3/2}\rangle$ level (emitting an $866\,$nm photon simultaneously) with a probability of $6\%$, and to the other Zeeman level of $|S_{1/2},m=-1/2\rangle$ with a probability of $\frac{1}{3}\times(1-6\%)\approx31\%$ (see Fig.~\hyperref[FIG2]{2(c)}). In addition, the ion will not be influenced by the $397\,$nm excitation pulse if it has already decayed to $|D_{3/2}\rangle$ state. Thus the ion returns to the initial state with a high probability after an unsuccessful excitation and will not be excited again after a successful excitation. Therefore, we can iteratively excite the communication ion by successive excitation pulses until all the population in the initial state has been removed. This scheme effectively improves the branching ratio of the $866\,$nm transition and we call this method `pushing the branching ratio to unity' in a vivid way. During the process of multiple excitation, a time-bin train of $866\,$nm photonic pulses are emitted and collected (see Fig.~\hyperref[FIG3]{3}), which  can be employed to enhance the efficiency of ion-photon entanglement generation.

After a series of four successive excitation pulses, most of the population in $|S_{1/2},m_S=+1/2\rangle$ is pumped to $|S_{1/2},m_S=-1/2\rangle$ and $|D_{3/2}\rangle$ level, and then the excitation probability becomes low. As most of the population is accumulated in $|S_{1/2},m_S=-1/2\rangle$ level, we can pump the ion back to the initial state via a $397\,$nm $\sigma^+$ transition (see Fig.~\hyperref[FIG2]{2(c)}). Note that unlike the initial pumping, here $866\,$nm laser is not applied to protect the state in $|D_{3/2}\rangle$. Iteratively, other series of excitation pulses and intermediate optical pumpings are applied, until most of the population are in $|D_{3/2}\rangle$ level. In this way, we can generate a time-bin train of $866\,$nm photonic pulses, which is temporally separable and can be employed to enhance the efficiency of ion-photon entanglement generation. Note that as the ion can also decay to $|D_{3/2}\rangle$ level in the intermediate optical pumping, thus not all the ion population in $|D_{3/2}\rangle$ level is related to photon emission event in the ion-photon entangling attempts. However, this does not influence the quality of ion-photon entanglement as the $866\,$nm photon emitted in the optical pumping will not be collected. In addition, the already created ion-photon entanglement will not be influenced by the following intermediate pumping as the ion is already in $|D_{3/2}\rangle$ state in this case. Therefore, the intermediate pumping will not induce false heralding or ruin the established ion-photon entangled state. Ideally, the branching ratio of the $866\,$nm emission in the ion-photon entangling can be `improved' to $54.4\%$ considering the optical pumping processes, and the multiplexing enhancement factor saturates with $N=10$ modes approximately. In the future, we can shelve the state in $|D_{3/2}\rangle$ to $|D_{5/2}\rangle$ and apply $866\,$nm laser in the intermediate optical pumping, by which we can achieve an effective branching ratio of $100\%$. Currently, we are limited by the laser power to realize  this fast shelving. The detailed analysis on this issue can be found in the Appendix \ref{app:ebr}.

Another scheme to realize multiplexing enhancement is ion shuttling. Through this scheme, one can successively generate temporally multiplexed ion-photon entanglements via shuttling different ions to the interaction region of the optical interface (either a high-NA objective or a high-finesse cavity~\cite{lanyon multiplexing,haffner multiplexing}). In our experiment, the time cost for ion shuttling is $\Delta t=25\,\mu$s (see Appendix \ref{app:is}).

\subsection{Multiplexing-enhanced ion-photon entanglement in three different cases}

In this section, we demonstrate the realization of multiplexing-enhanced ion-photon entanglement via the two multiplexing schemes mentioned above, at different fiber lengths of $3\,$m, $1\,$km, and $12\,$km.

\begin{figure}[htbp]
    \centering
    \includegraphics[width=0.48\textwidth]{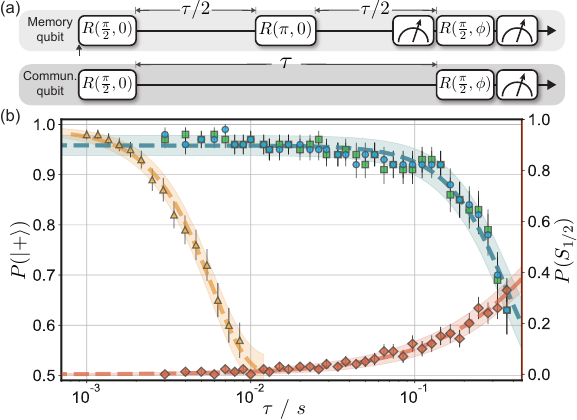}
    \caption{
    Coherence time and crosstalk.
    (a) Protocol to measure the coherence time of communication qubit and memory qubit.
    We prepared the memory qubit in $|+_M\rangle=\frac{|\downarrow_M\rangle+|\uparrow_M\rangle}{\sqrt{2}}$ (Fig.~\hyperref[FIG2]{2(b)}), and apply a Ramsey measurement with a spin echo in the middle. For the communication qubit, we prepare the communication qubit in $|+_C\rangle=\frac{|\downarrow_C\rangle+|\uparrow_C\rangle}{\sqrt{2}}$ and apply a Ramsey measurement without spin echo.
    (b) Measurement results.
    For the communication qubit, a fitting of Gaussian decay yields a coherence time of $5.8\pm 0.1\,$ms (yellow triangle).
    The green square and the fitted curve show the fidelity decay in the memory qubit with all the noisy operations on the communication qubits applied. The extracted coherence time is $366\pm 11\,$ms.
    The blue circle is the memory qubit fidelity in the case of no operations on the communication qubits, and the fitted coherence time is $368\pm9\,$ms. The coherence time of memory qubit with or without operations on communication qubits cannot be faithfully distinguished considering the statistical error. The orange diamonds characterize the spontaneous decay of the memory qubit after different storage times. The fitted lifetime of $|D_{5/2}\rangle$ is $958\,$ms. The small discrepancy between our measurement result and the theoretical value ($\sim1\,$s) can be attributed to the leaked $854\,$nm laser.
    All the coherence time is fitted by a Gaussian decay $F=a\,e^{-(x/\tau)^2}+1/2$.
    }
    \label{FIG4}
\end{figure}

\begin{figure*}[htbp]
    \centering
    \includegraphics[width=0.95\textwidth]{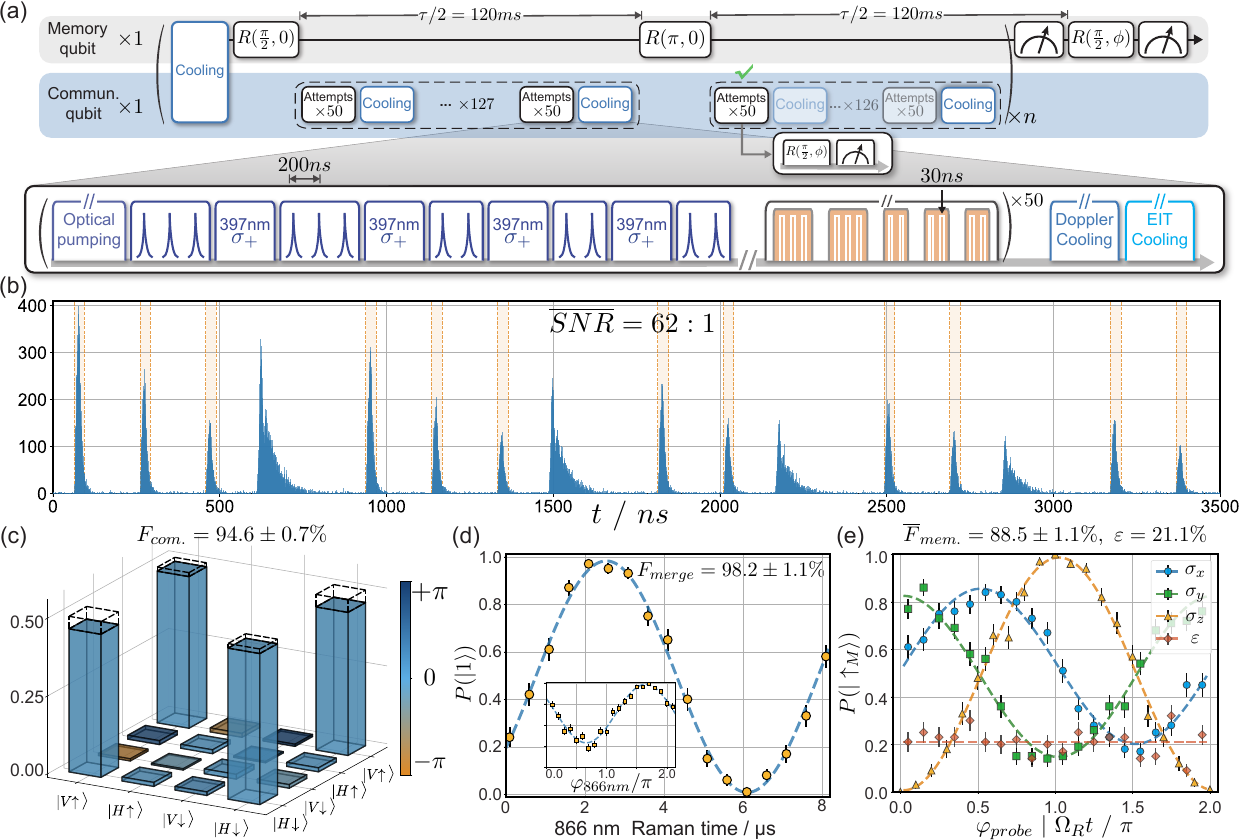}
    \caption{
        Experimental results of the multiplexing-enhanced ion photon entanglement in $1\,$km case.
        (a) Experimental protocol. Unlike the $3\,$m case, we increase the time interval between successive excitation pulses to around $200\,$ns. We apply a series of $3-3-2-2-2$ excitation pulses and $4$ optical pumpings in between.
        (b) Histogram for photon detection events. The time interval between different modes is set to $200\,$ns. Four intermediate optical pumpings are inserted between excitation pulses.
        (c) Reconstructed density matrix of the ion-photon entanglement. The measured fidelity is $94.6\pm0.7\%$.
        (d) The fidelity of the merging operation is $98.2\pm1.1\%$.
        (e) The memory qubit measurements in six mutually unbiased bases (MUBs).
        The survival rate of memory qubit is about $79\%$, and the averaged fidelity over six MUBs is $88.5\pm 1.1\%$ after $240\,$ms storage.
    }
    \label{FIG5}
\end{figure*}

For the $3\,$m case, the round-trip travel time for heralding an ion-photon entanglement is $\frac{2L}{c}=30\,$ns and the averaged overhead $T_{\text{ovh}}$ is below $1\,\mu$s in our experiment. In this situation of small overhead and short fiber, a small $\Delta t$ is necessary to achieve a high multiplexing enhancement factor. Therefore, it is not suitable to use the ion-shuttling method in this case as the $25\,\mu$s shuttling time is much higher than the total overhead $\frac{2L}{c}+T_{\text{ovh}}\approx1\,\mu$s. Here we use a $2$-ion chain (one for communication qubit and the other for memory qubit) and employ the multiple excitation method to accelerate the generation of heralded ion-photon entanglement. As the enhancement factor of this method has an upper limit of $\frac{54.4\%}{6\%}\approx9$ (all the population is pushed to $|D_{3/2}\rangle$ level), which means the enhancement factor already saturates roughly at $N\approx9$. Therefore, we need to reduce $\Delta t$ to a value as small as possible, in order to guarantee $N\Delta t\approx 9\Delta t\approx\frac{2L}{c}+T_{\text{ovh}}$ and achieve a high enhancement factor (see the analysis in Section \ref{text:efc}). Therefore, here we use the fastest multiple excitation scheme in which we apply a series of four consecutive pulses from a mode-locked pulse laser to excite the ion. The time interval between consecutive pulses is $13\,$ns which is the reciprocal of the laser repetition rate at $76\,$MHz. After $4$ consecutive pulses, we pump the ion back to the initial state by a $397\,$nm pulse (intermediate pumping). Another set of $4$ consecutive pulses is applied after the intermediate optical pumping. A total number of $8$ excitations are applied to generate $8$ photonic time-bin modes (see Fig.~\hyperref[FIG3]{3}). Upon any of the $8$ time-bin modes is detected by the photon detector, our control system can identify which mode is excited via the arrival time of the photon detection signal. Followed by dynamical adjustment of the ion phase, the finally yielded ion-photon entangled state $\frac{|\uparrow\rangle|H\rangle+e^{i\phi}|\downarrow\rangle|V\rangle}{\sqrt{2}}$ always has the same phase $\phi$ no matter which mode is actually excited. The heralded ion-photon entanglement has a fidelity of $96.8\pm0.4\%$ to the closest maximally entangled state. The total time cost of the $8$ excitation pulses and the intermediate pumping is about $340\,$ns, and the entangling rate is enhanced by a factor of $M=3.4$ compared to the single-mode case. The entanglement generation rate in this case is about $263\,$s$^{-1}$. As the time interval between the excitation pulses $13\,$ns is comparable to the spontaneous decay time of $|P_{1/2}\rangle$ at $7\,$ns, it is possible that the ion is still in the excited state when the next excitation pulse arrives. However, this only induces a slight efficiency loss but does not influence the quality of the heralded entanglement, which is discussed in Appendix \ref{app:iie}. The memory qubit fidelity averaged over six mutually unbiased bases (MUBs)~\cite{huangyy, lai} is measured to be $93.3\pm0.9\%$ after $100\,$ms storage.

As a functional quantum network node should be capable of storing quantum information during the ion-photon entangling attempts, we use the dual-type scheme to encode quantum information in a memory qubit in $|D_{5/2}\rangle$ level for this task~\cite{huangyy,lai,omg}. It is shown that the coherence time of the quantum information in the memory ion is $366\pm 11\,$ms and the crosstalk of operations on communication qubit has negligible influence on the memory qubit, as shown in Fig.~\hyperref[FIG4]{4}. In this case, the memory coherence time is roughly $100$ times longer than the ion-photon entanglement generation time of $\frac{1}{263\text{s}^{-1}}=3.8\,$ms.

For the $1\,$km case, the total overhead $\frac{2L}{c}+T_{\text{ovh}}$ is slightly higher than $10\,\mu$s. As the total overhead is still small compared to the $25\,\mu$s switching time of ion shuttling, we still use the multiple excitation method as in the $3\,$m case, and only one communication qubit is used. However, now we can set the time interval $\Delta t$ to a larger value of around $200\,$ns to avoid the efficiency loss in the $13\,$ns interval case, as the population in the excited state is negligible after $200\,$ns. The pulse sequence in this case is shown in Fig.~\hyperref[FIG5]{5(a)}, in which totally $12$ excitation pulses are applied to excite ion-photon entanglement in a total duration of $3.3\,\mu$s, including the four optical pumping in between. In this case, the generation rate of heralded ion-photon entanglement is $40\,$s$^{-1}$, enhanced by a factor of $5.1$ via multiplexing. The fidelity of the yielded ion-photon entanglement after feedforward on heralding signal is measured to be $94.6\pm 0.7\%$, as shown in Fig.~\hyperref[FIG5]{5(c)}. The memory fidelity averaged over six MUBs is $88.5\pm 1.1\%$ after $240\,$ms storage.

\begin{figure*}[htbp]
    \centering
    \includegraphics[width=0.95\textwidth]{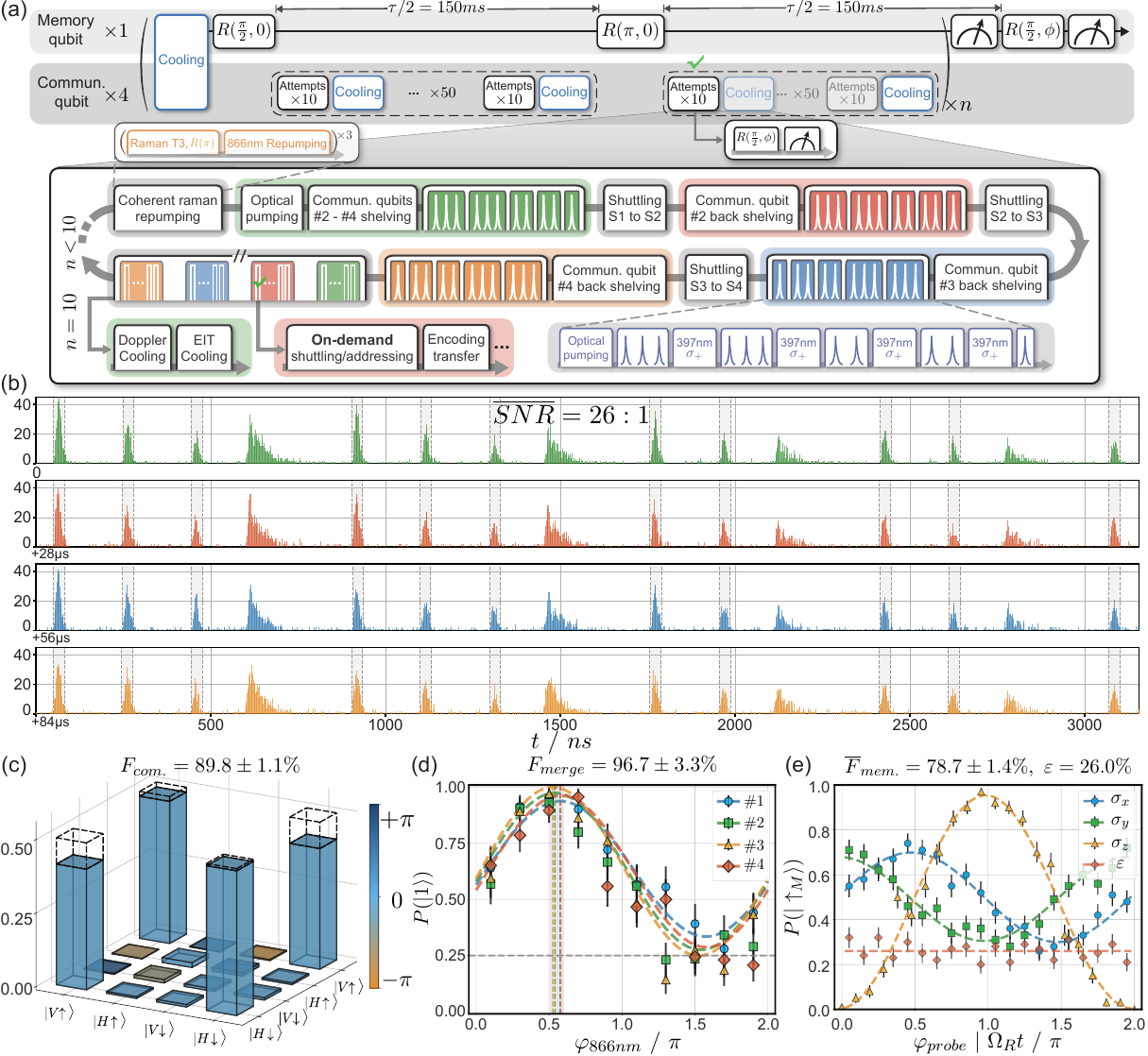}
    \caption{
        Experimental results of the multiplexing-enhanced ion photon entanglement in $12\,$km case.
        (a) Experimental protocol. In this case, we use a hybrid multiplexing method which combines the multiple excitation scheme used in the cases of $3\,$m and $1\,$km, and the ion shuttling scheme. In one round of excitation, each of the $4$ communication quibt is excited by $11$ times and emits in $11$ time-bin photonic modes via multiple excitation scheme, and totally $44$ modes are emitted after all the $4$ communication qubits are excited via ion shuttling.
        A $650\,\mu$s sympathetic cooling is applied after $10$ rounds of excitation.
        In this protocol, we apply on-demand shuttling and addressing for the measurement of communication qubits and memory qubit.
        (b) The histogram of the $44$ photonic modes.
        (c) The measured ion-photon entanglement fidelity is $89.8\pm1.1\%$.
        (d) The fidelity of merging operation on the $4$ communication qubits. Here we fix the pulse area to $2/3\,\pi$ and scan the phase for the merging pulse.
        This merging transition is sensitive to magnetic field, and we can use this property to check the magnetic field inhomogeneity on the $4$ communication qubits (see Appendix~\ref{app:mfh} for details).
        This measurement yields a fidelity of $96.7\pm3.3\%$.
        (e) The memory qubit measurement in six mutually unbiased bases (MUBs).
        The survival rate of memory qubit is about $74\%$, and the averaged fidelity over six MUBs is $78.7\pm 1.4\%$ after $300\,$ms storage.
    }
    \label{FIG6}
\end{figure*}

For the case in which the fiber length is $12\,$km, the total overhead $\frac{2L}{c}+T_{\text{ovh}}$ is larger than $120\,\mu$s, which is already longer than the switching time of ion shuttling. In this case, we accelerate the remote entangling process via a hybrid of both multiplexing methods mentioned in Section~\ref{text:mem}, as shown in Fig.~\hyperref[FIG6]{6}. Unlike the $3\,$m and $1\,$km cases, here we increase the number of communication qubits from $1$ to $4$ as shown in Fig.~\hyperref[FIG2]{2(a)}, and convert the $866\,$nm photon to $1558\,$nm via DFG~\cite{lai}. We first use the similar time-bin multiplexing method as in the $1\,$km case to generate $11$ time-bin pulses from ion $1$. Then we shuttle the ion chain to move ion $2$ to the focus of the objective and generate another $11$ time-bin pulses from ion $2$, and later ion $3$ and ion $4$. For the communication qubits not in the focus of the objective, their population in $|D_{3/2}\rangle$ level are shelved to $|D_{5/2}\rangle$ during the excitation of other communication qubits to avoid the crosstalk influence. After all the excitations, we shuttle the ion chain back to the initial position. The total duration of the ion-photon entangling attempts lasts for $87\,\mu$s with totally $44$ time-bin modes emitted. The remote entanglement generation rate is $4.28\,$s$^{-1}$ and the multiplexing enhancement factor is $15.6$ compared with the single-mode case. Here we still use the arrival time of photon detection events for feedforward and yield the same ion-photon entanglement no matter which of the $44$ modes is excited, and the final heralded ion-photon entanglement has a fidelity of $89.8\pm 1.1\%$. The dephasing time of the memory qubit is $366\pm 11\,$ms, which is already longer than the expected entanglement generation time about $\frac{1}{4.28\,\text{s}^{-1}}=234\,$ms. Thus with multiplexing enhancement we can achieve a ratio of the remote entangling rate to the memory decoherence rate (also known as `link efficiency'~\cite{hansondelivery,duan pra}) of $\frac{1}{234\,\text{ms}}/\frac{1}{366\,\text{ms}}\times74\% =1.16>1$ ($74\%$ is the survival rate of memory qubit, see Fig.~\hyperref[FIG6]{6} and the analysis below), which is an important threshold for the scale-up of a quantum network and sets the record for metropolitan-scale fiber so far.

As the memory qubit is encoded in $|D_{5/2}\rangle$ state which has a finite lifetime of $958\,$ms in our experiment (see Fig.~\hyperref[FIG4]{4(b)}), the qubit stored in the memory qubit will suffer from considerable spontaneous emission error when the protocol time is close to the lifetime. However, this error can be detected via a mid-circuit measurement, thus it will not influence the quality of the quantum storage but only reduces the success probability of the task~\cite{lai}. In this experiment, we still use this mid-circuit measurement scheme to detect and avoid the influence of the spontaneous emission error. The spontaneous decay error happens with a probability of $11\%$, $21\%$, and $26\%$ in the $3\,$m, $1\,$km, and $12\,$km cases, respectively (see Figs.~\hyperref[FIG3]{3(e)}, \hyperref[FIG5]{5(e)}, and \hyperref[FIG6]{6(e)}).\\

\section{Discussion}

In summary, we have experimentally realized a highly-multiplexed trapped-ion quantum network node, which is capable of delivering heralded ion-photon entanglement with an entangling rate higher than the memory decoherence rate over a metropolitan-scale fiber. The remarkable enhancement factor of about $16$-fold is achieved via the hybrid of a novel multiple excitation scheme which is demonstrated for the first time in this work and the ion shuttling scheme. Besides, the stored quantum information in the memory qubit is protected from the dissipative operations on the communication qubit via dual-type encoding, and the spontaneous emission error in the memory qubit can be detected via a mid-circuit meansurement. In the future, the generation rate of the heralded ion-photon entanglement can be further improved by optimizing the efficiency of wavelength conversion from $12\%$ in this experiment to the current best $57\%$~\cite{bock}. Currently, the ion shuttling speed is limited by the bandwidth of the low-pass filter for the trap electrodes, and future improvement to a shuttling time of less than $3\,\mu$s is possible by replacing the filter. The entanglement generation rate can also be largely enhanced if high-finesse cavity is applied in the future~\cite{866 cavity}. With all these improvements, we believe a multi-node, metropolitan-scale trapped-ion quantum network  could be realized in the future based on the multiplexed quantum network node demonstrated in this work.

\section{Acknowledgements}
This work is supported by Innovation Program for Quantum Science and
Technology (No.2021ZD0301604, No.2021ZD0301102), the Tsinghua University
Initiative Scientific Research Program and the Ministry of
Education of China through its fund to the IIIS. Y.F.P. acknowledges
support from the Dushi Program
from Tsinghua University.

\appendix

\section{Multiplexing enhancement with inhomogeneous success probability}\label{app:dmec}

In this section we calculate the enhancement factor of a multiplexing scheme when the success probability of different time-bin modes are different. Here we assume the success probability in the single-mode case is $p_0$, and the success probabilities of the $N$ time-bin modes are $p_0, p_1, \dots, p_N$. Then, the enhancement factor is:
\begin{widetext}
\begin{equation}\label{mer}
M'=R_{\text{mul}}/R_0=\frac{\sum_{i=0}^{N}p_i}{\frac{2L}{c}+T_{\text{ovh}}+N \Delta t}/\frac{p_0}{\frac{2L}{c}+T_{\text{ovh}}+ \Delta t}=M\frac{\sum_{i=0}^{N}p_i}{N}/p_0
\end{equation}
\end{widetext}
where $R_{\text{mul}}$ is the success rate with multiplexing enhancement and $R_0$ is the success rate in single-mode case. Eq.~\eqref{mer} shows that, in the case that the success probability is constant ($p_i=p_0$ for each $i\in [1,N]$), the multiplexing enhancement factor $M'$ will be equal to $M$ in Eq.~\eqref{M}.

\section{Effective branching ratio}\label{app:ebr}

\begin{figure}[htbp]
    \centering
    \hspace{-0.4cm}
    \includegraphics[width=0.45\textwidth]{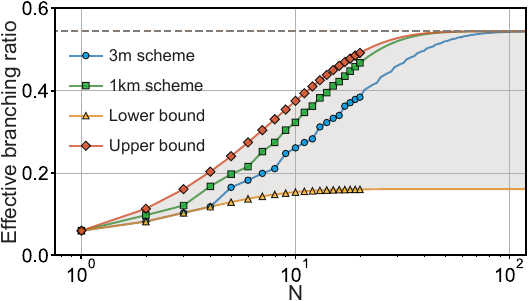}
    \caption{
        Effective branching ratio as a function of excitation number. Here $N$ is the number of excitation pulses.
        The upper bound is achieved when every excitation is followed by an intermediate pumping. The $3\,$m$/$$1\,$km case follows the schemes with $3\,$m$/$$1\,$km fiber in the main text. The lower bound is achieved without intermediate optical pumping.
    }
    \label{FIG7}
\end{figure}

In this section, we discuss the effective branching ratio via the multiple excitation scheme.
The entire attempting process including excitation pulses and intermediate pumpings can be modeled as a Markov decision process,
in which each operation is defined by a corresponding transition matrix.
The shaded region in Fig.~\hyperref[FIG7]{7} represents the region of all possible strategies.
The lower bound corresponds to the strategy only containing $397\,$nm excitation pulses without any intermediate pumping.
By applying many successive $397\,$nm excitation pulses of $\sigma^-$ polarization,
$p_{br,S}\times\,p_{cg,-1/2\rightarrow-1/2}/(p_{br,S}\times\,p_{cg,-1/2\rightarrow -1/2}+p_{br,D}) = 83.9\%$ of the population is transferred to $|S_{1/2},m=-1/2\rangle$, where $p_{br,i}$ and $p_{cg,-1/2\rightarrow-1/2}$ are defined as the
branching ratio from $|P\rangle$ to $|i\rangle$ ($i=S_{1/2}$ or $D_{3/2}$) and square of CG coefficient from $|P_{1/2},m=-1/2\rangle$ to $|S_{1/2},m=-1/2\rangle$, respectively. In this case, most population of the ion will be accumulated in the $|S_{1/2},m=-1/2\rangle$ state and an effective `branching ratio' of $16.1\%$ of the $866\,$nm transition can be achieved.

We can further improve the effective branching ratio by inserting intermediate optical pumpings between the excitation pulses (Fig.~\hyperref[FIG3]{3}, \hyperref[FIG5]{5} and \hyperref[FIG6]{6}), as the unwanted population accumulated in $|S_{1/2},m=-1/2\rangle$ can be removed.
As illustrated in the simulation based on master equations (Fig.~\hyperref[FIG7]{7}), all our schemes eventually converge to a single maximum effective branching ratio of ${BR}_{max} = 54.4\%$ as the number of modes increases. The reason that the effective branching ratio cannot reach $100\%$ is that the ion can also decay to $|D_{3/2}\rangle$ state during the intermediate pumping. Thus not all the ion population in the $|D_{3/2}\rangle$ level is related to photon emission events in the ion-photon entangling attempts.

In addition, we can improve the effective branching ratio to $100\%$ with some modifications in the intermediate pumping in the future. We can shelve the population in $|D_{3/2}\rangle$ level to the memory qubit subspace $|D_{5/2}\rangle$ before the intermediate pumping and then convert it back to $|D_{3/2}\rangle$ after the pumping. Via this shelving, we can shine $866\,$nm laser in the intermediate pumping without influencing the already established ion-photon entangled state in this process. Unlike the intermediate pumping used in this experiment which does not include any $866\,$nm laser, in this case all the population in $|D_{3/2}\rangle$ are generated in the ion photon entangling process. Thus we can improve the effective branching ratio to $100\%$ via this method. The reason for not applying this method in our current experiment is due to the low power of the Raman laser in our current setup. Under current laser power, the shelving operation is too slow which would significantly limit the speed of entangling operations if the shelving is applied. In the future, if the power of Raman laser can be significantly improved, we can achieve an effective branching ratio closer to unity and further improve the distribution rate of ion-photon entanglement.

\section{Ion shuttling}\label{app:is}

A home-made PCB was developed to supply stable DC voltages for the DC electrodes.
The board is based on a 32-channel, 16-bit AD5372 chip with specially designed low-pass filters.
To perform shuttling operations for $N$ ions, we utilize ADG5409 as the analog signal multiplexer.
Each ADG5409 routes one of four differential inputs to a common differential output,
as selected by the 2-bit binary address lines. Each chip comprises two such channel multiplexers.
Consequently, the system requires $4N$ analog channels of the AD5372,
$\log_2N$ digital control channels and $2\log_2N$ ADG5409 chips for $N$ ions multiplexing scheme.
An additional $200\,$kHz low-pass filter is applied to the outputs D1-D4
to ensure a low noise DC voltage for the shuttling operation.

\begin{figure}[htbp]
    \centering
    \includegraphics[width=0.48\textwidth]{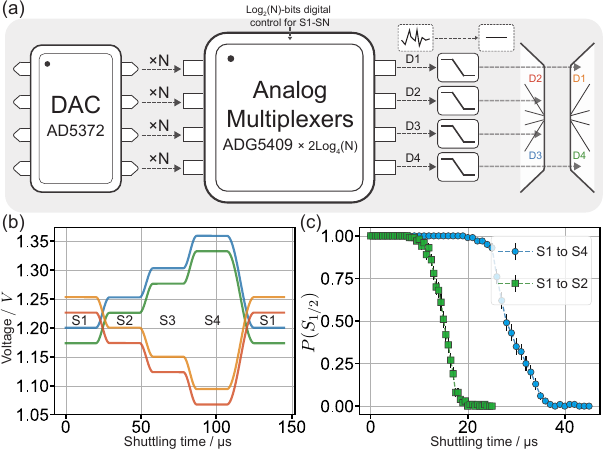}
    \caption{
    Setup for ion shuttling and the performance.
    (a) The architecture of the circuit. The circuit is composed of a DAC, low-pass filters, and analog multiplexers for shuttling.
    (b) Simulation of the electrodes' voltage in the shuttling. Different colors correspond to different blade segments as shown in (a). (c) Measurement of the ion shuttling speed using a $729\,$nm addressing beam. Here we test the moving speed by shuttling communication qubit 1 to the position of communication qubit 2 (green square), as well as shuttling communication qubit 1 to communication qubit 4 (blue circle).
    }
    \label{FIG8}
\end{figure}

A simulation (Fig.~\hyperref[FIG8]{8(b)}) is performed to model the time evolution of the voltage on the blades during the shuttling sequence from state S1 (communication qubit 1 is in the focus region of the objective) to S4 (communication qubit 4 is in the focus region of the objective) and then back to S1.
To directly measure the shuttling time, we used the $729\,$nm addressed laser to shelve an communication qubit to the $|D_{5/2}\rangle$ state at a fixed position in the ion chain.
Observing the dynamics of this shelving process allows us to roughly estimate the ion shuttling time.
In the current $12\,$km scheme, a shuttling distance of $8.5\,\mu$m requires $20\,\mu$s as depicted in Fig.~\hyperref[FIG8]{8(c)}.

\section{Multiplexing enhancement of heralded ion-ion entanglement}\label{app:iie}

\begin{figure}[htbp]
    \centering
    \includegraphics[width=0.48\textwidth]{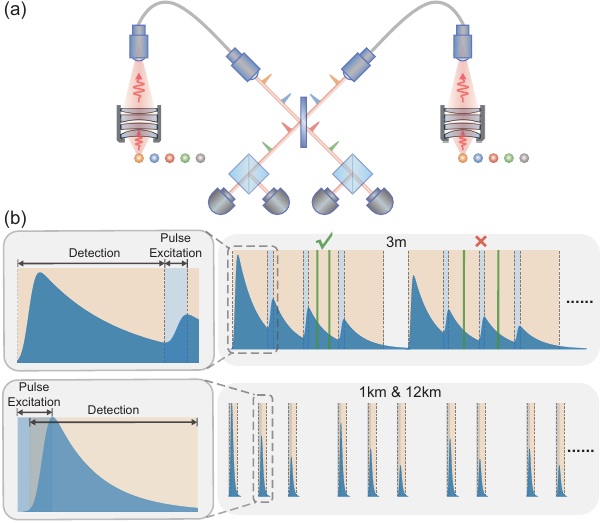}
    \caption{
    Bell state measurement and time windows for multiplexed ion-ion entanglement.
    (a) The Bell state measurement setup in the ion-ion entanglement.
    The apparatus comprises a 50:50 beam splitter, two polarization beam splitters, and four single photon detectors.
    By selecting corresponding coincidence patterns, two distinct entangled states $|\psi^\pm \rangle = |01\rangle \pm |10\rangle$ can be generated.
    (b) The time windows for excitation and detection in different cases.
    }
    \label{FIG9}
\end{figure}

In this section, we analyze how to apply the multiplexing method demonstrated in this work to accelerate the heralded ion-ion entanglement in the future. First, we explain why the time-bin modes generated in the method of `pushing the branching ratio' to unity can guarantee perfect ion-ion entanglement, especially in the $3\,$m case when the time interval between successive excitation pulses is comparable to the lifetime of the excited state. Here we assume the ion-ion entanglement is heralded via two-photon coincidence in photonic Bell state measurement (BSM), as shown in Fig.~\hyperref[FIG9]{9(a)}. We use $N$ detection windows to herald the ion-ion entanglement, the entangling attempt is successful when in one of the $N$ windows a two-photon coincidence is detected ($N$ is the number of excitation pulses).

The quality of the ion-ion entanglement is determined by the indistinguishability of the photons emitted from both sides. Thus the photons from both ions should have the same frequency and temporal shape of wavepacket. The matching of frequency can be implemented by applying the same magnetic field to both setups, and the time difference of wavepacket can be aligned via compensating the optical path or adjusting the time of the excitation pulse precisely. As both of the two remote ions are in free space, the photonic wavepackets are both in a shape of exponential decay with the same characteristic time. Thus the perfect indistinguishability in photons from both sides can be guaranteed ~\cite{oxford_qkd, monroe}. Compared with the single-mode case or multiplexing with only ion shuttling~\cite{lanyon multiplexing, haffner multiplexing}, the main difference in the multiple excitation scheme lies in that the ion is exposed to a series of successive excitation pulses. However, this is not a problem. In this method, once one of the ions decays to $|D_{3/2}\rangle$ state, the excitation process is effectively terminated, as the following excitation attempts or intermediate pumpings will never yield a two-photon coincidence because the ion in $|D_{3/2}\rangle$ state is already not affected by the later excitation pulses. This guarantees two results: 1) the two-photon coincidence can only happen once both ions are excited to the $|P_{1/2},m=-1/2\rangle$ state by the excitation pulse and decay to $|D_{3/2},m=-3/2,-1/2,+1/2\rangle$ state in the same detection window (see Fig.~\hyperref[FIG9]{9(b)}), in which case the ion-ion entanglement is successfully generated, and 2) once the ion-ion entanglement is heralded, the following excitation pulses will not influence the state of both ions. There exist cases that both ions decay back to $|S_{1/2}\rangle$ state in an excitation pulse or intermediate pumping, in this case no coincidence will be detected and the excitation process will continue.

In short, there can be $3$ cases in each excitation attempt: 1) Both ions emit a $866\,$nm photon. In this case, an ion-ion entanglement is successfully heralded and the ion state will not be influenced by the following excitation pulses and intermediate pumpings. 2) Neither ion emits a $866\,$nm photon. In this case the excitation attempts will continue. 3) One ion emits a $866\,$nm photon but the other does not. In this case, the excitation process is effectively ended, and a two-photon coincidence will never be detected. Therefore, the already established ion-ion entanglement will not be influenced by the following excitation pulses or intermediate pumpings, and there is no false heralding events. These two properties guarantees the quality of ion-ion entanglement. Another point worth noting is, in the two-photon BSM, the relative difference in the photon emission probabilities from each ion does not influence the quality of the heralded ion-ion entanglement, but only influences the success probability. This guarantees that the heralded ion-ion entanglement by the two-photon BSM is in perfect quality despite efficiency imbalance.

\begin{figure*}[htbp]
    \centering
    \includegraphics[width=0.8\textwidth]{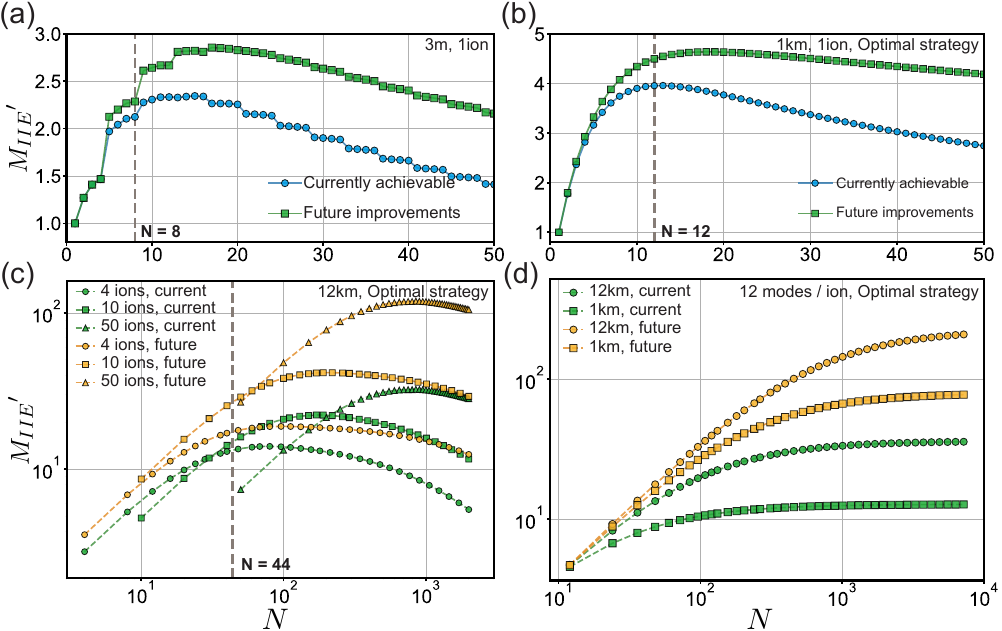}
    \caption{
        Simulations of multiplexing enhancement for ion-ion entanglement.
        (a) and (b) illustrate the simulations of multiplexing enhancement for ion-ion entanglement at distances of $3\,$m and $1\,$km with only the multiple excitation scheme, respectively. Here $N$ is the number of excitations.
        The blue curves and data points represent the currently achievable conditions as shown in this work.
        In the $1\,$km scenario, we assume the optimal strategy is employed (here we assume each excitation pulse is followed by an intermediate pumping, see Appendix \ref{app:ebr}).
        For future improvements, the maximum enhancement achievable is represented by the green curve and data points. In this case we assume $\Delta t=52\,$ns is used.
        (c) The simulation of multiplexing enhancement in $12\,$km case. The conditions are similar with (b) (we also assume $T_{\text{ovh}}=50\,\mu$s). The only difference is that we assume the shuttling time is $3\,\mu$s in the curves and data points regarding future improvements. Here $N$ is the number of total modes used (i.e., number of temporal modes per communication qubit $\times$ number of communication qubits).
        (d) The multiplexing enhancement of ion-ion entanglement with varying ion numbers. Here $N$ is still the number of total modes.  Here we fix the number of excitations in each ion to $12$ but vary the number of ions, which is different from (a), (b), and (c).
    }
    \label{FIG10}
\end{figure*}

In the $3\,$m case, the time interval $13\,$ns between successive excitation pulses is comparable to the excited state lifetime $7\,$ns. This will cause a situation that the ion is still in the excited state when the next excitation pulse arrives, and the ion will be flipped back to $|S_{1/2}\rangle$ state by this $6\,$ps $397\,$nm $\pi$ pulse. However, this does not influence the quality of the heralded ion-ion entanglement but only influences the success probability as discussed in the above paragraphs. As the $N$ detection windows do not overlap with the excitation pulses, in each detection window, the photons from both sides have the same wavepacket shape, i.e. exponential decay but not decay to zero (see Fig.~\hyperref[FIG9]{9(b)}). Thus the wavepacket shapes of both photons are still identical. The only influence is that the efficiency drops a little bit as the tail of the exponential decay is cut out off the detection window, compared to the $1\,$km and $12\,$km cases.

We simulate the factor of multiplexing enhancement in the ion-ion entanglement via the scheme described above (see Fig.~\hyperref[FIG10]{10}). The simulation is done in the realistic case which is currently achievable and in an ideal case  with future improvements. It is shown that the method `pushing the branching ratio to unity' can guarantee roughly a $5$-fold enhancement in efficiency compared to the single-mode case. However, the efficiency is still lower than a branching ratio of $1$. This can be improved by applying more ions via the shuttling method and faster shuttling speed in the future.

\section{Impact of magnetic field gradient on scalability}\label{app:mfh}

\begin{figure}[h]
    \centering
    \includegraphics[width=0.47\textwidth]{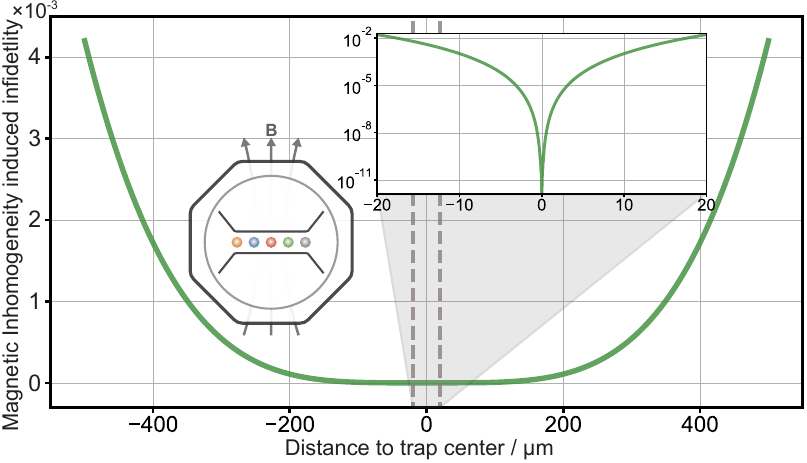}
    \caption{
        Infidelity induced by the magnetic field inhomogeneity.
        Numerical simulation of the infidelity induced by magnetic field inhomogeneity in the $12\,$km  scheme within a 1 mm range in the ion trap.
        The inset figure simulates the infidelity induced by magnetic field inhomogeneity after storing for $300\,$ms. The magnetic field inhomogeneity is simulated via finite element analysis based on the geometry of the trap.
    }
    \label{FIG11}
\end{figure}

\begin{figure*}[htbp]
    \centering
    \includegraphics[width=0.8\textwidth]{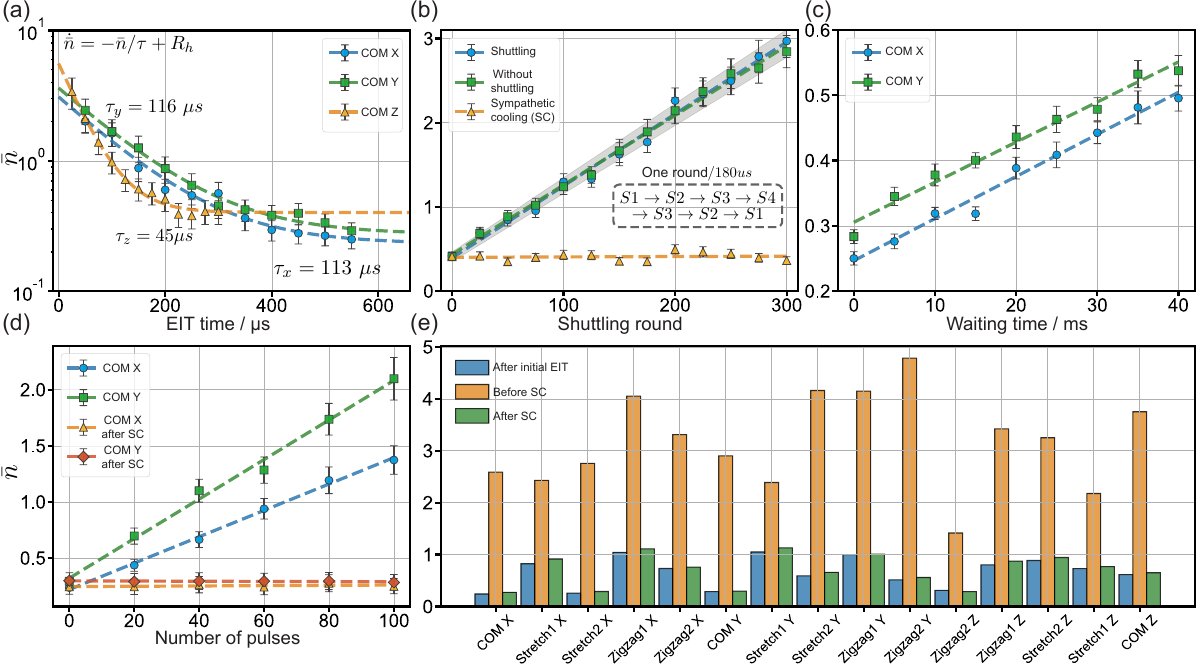}
    \caption{
        Measurements of heating and cooling effects.
        (a) $1/e$ cooling times $\{\tau_z,\tau_x,\tau_y\}=\{45\pm 1,113\pm 8,116\pm 6\}\,\mu$s
        for all three COM modes with average cooling limitations of $\{0.40\pm 0.02,0.24\pm 0.22,0.27\pm 0.10\}$.
        (b) The phonon number of the axial COM mode under different situations. The orange triangles are the measured phonon number with sympathetic cooling on. The blue circles are measured with ion shuttling and the green squares are measured without ion shuttling. No cooling is applied in blue circles or green squares. We can see that the shuttling has negligible heating in our case, as there's no obvious increase in the phonon number after $300$ rounds of shuttling, compared to the case of no shuttling.
        (c) The background heating in the radial COM modes. The fitted results in (b) and (c) yield a background heating rate of $\{46\pm 1, 6.5\pm 0.3,6.1\pm 0.3\}\,s^{-1}$.
        (d) The heating effects induced by the ion-photon entangling attempts in COM X and COM Y modes. Here we perform different numbers of excitation pulses after the initial cooling, followed by a sympathetic cooling. The green and blue data points are measured before the sympathetic cooling. The yellow and orange data points are measured after the sympathetic cooling. The measurements yield a  heating of $\{1.18(3)\times 10^{-2},1.76(8)\times 10^{-2}\}$ phonons for each excitation pulse in these two radial modes.
        (e) The measured phonon numbers for all the $15$ modes after the initial cooling, after the entangling attempts, and after the sympathetic cooling, in the protocol of the $12\,$km scheme.
    }
    \label{FIG12}
\end{figure*}

Implementing a multiplexing scheme requires
not only the efficient distribution of remote entanglement
but also the capability to make all the yielded ion-photon entanglement the same for subsequent computation and processing.
Thus a critical requirement is to guarantee the consistency of the ion phase when the heralding event happens on different communication qubits.
In polarization-encoded schemes, communication qubits are encoded on Zeeman sublevels, making them inherently sensitive to magnetic field.
As a result, the homogeneity of the magnetic field is very important when the ion number is high.
Although individual addressing and phase feedforward control have been realized, maintaining magnetic field homogeneity over a larger spatial region can significantly reduce the complexity of scaling up the system.

Here we build a Helmholtz coil using specially arranged cobalt magnets to generate a magnetic field of $4.16(5)\,$G.
We evaluate the impact of magnetic field inhomogeneity of our system from two perspectives.
Firstly, the phase accumulation in the entangling attempts should be the same for each ion in the communication qubit chain. During the ion shuttling, different communication qubits experience different magnetic fields due to the spatial separation between them. Therefore, the magnetic field inhomogeneity determines the phase inhomogeneity of the communication qubits after one round of excitation.
Using finite element analysis, we simulate the infidelity induced by magnetic field inhomogeneity in the $12\,$km scheme within a $1\,$mm range over the ion trap.
The result shows that, within a chain of $120\,$communication qubits, this infidelity is less than $5\times10^{-3}$ (Fig.~\hyperref[FIG11]{11}), which is sufficient to achieve the required mode numbers $N$ for saturated multiplexing enhancement demonstrated in Fig.~\hyperref[FIG10]{10(d)}.
Experimentally, we scan the phase of the $866\,$nm$/$$866\,$nm Raman merging rotation to confirm the magnetic field inhomogeneity is small (see Fig.~\hyperref[FIG6]{6(d)}). This result shows that the phases of the four communication qubits are within one standard deviation of their mean value,
noting that the $866\,$nm$/$$866\,$nm Raman merging operation is most sensitive to magnetic field variations among all operations in our experiment.

Secondly, for subsequent computational tasks, the phase discrepancies caused by magnetic field inhomogeneity in the memory qubit must remain sufficiently small.
The inset figure of Fig.~\hyperref[FIG11]{11} simulates the infidelity induced by magnetic field inhomogeneity after storage of $300\,$ms, which shows infidelity less than $10^{-2}$ within a range of $40\,\mu$m.

\section{Heating processes and cooling scheme}
\begin{figure*}[htbp]
    \centering
    \includegraphics[width=0.8\textwidth]{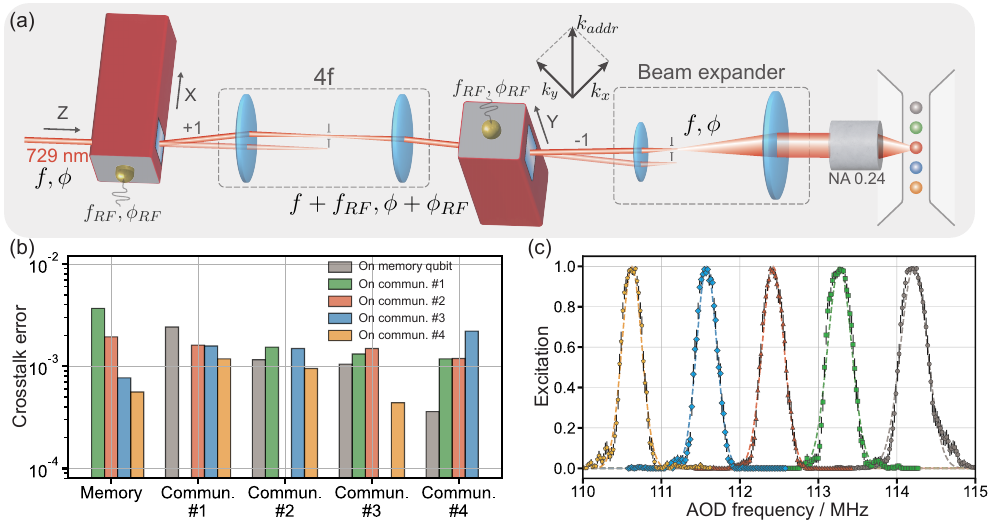}
    \caption{
        Setup and performance for the $729\,$nm addressing system.
        (a) Experimental setup for the addressed beam.
        (b) Crosstalk on neighboring ions when the spot is focused on different ions.
        (c) Excitation when scanning the addressed beam across the ion chain.
    }
    \label{FIG13}
\end{figure*}

We confine a crystal of five ions with three COM mode frequencies of $\{\omega_z,\omega_x,\omega_y\}=2\pi\times\{0.28,1.25,1.09\}\,$MHz.
EIT (Electromagnetically Induced Transparency) cooling is employed to cool the $4$ communication qubits directly and the memory qubit sympathetically.
By measuring cooling rates and investigating heating processes, we select appropriate durations and duty cycles for sympathetic cooling.
Efficient cooling is critical to achieve high fidelities for operations in all communication qubits and memory qubit.
We measure the background heating rates for all COM modes, heating in axial COM mode induced by shuttling, and the heating induced by $397\,$nm excitation pulses in radial COM modes under the situation of single ion for convenience (Fig.~\hyperref[FIG12]{12}).
In addition, average phonon numbers in $15\,$ motional modes before and after $10\,$ rounds of operations following the $12\,$km protocol ($440$ excitations in total) are measured, as well as after implementing sympathetic cooling.

\section{Individual addressing}

Fig.~\hyperref[FIG13]{13} illustrates the crossed AOD configuration.
Two acousto-optic deflectors (AODs) with principal axes perpendicular to each other are driven by a microwave source.
By combining the positive and negative first-order diffractions from each AOD,
we achieve the ability to scan the ion chain.

With an objective of NA$\,=0.24$, we can focus the beam to a spot with a diameter of $4.18\,$um.
The crosstalk error, defined as the ratio of the residual Rabi frequency on neighboring ions to that on the addressed ion,
is measured to be less than $4\times 10^{-3}$.

\end{document}